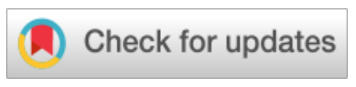

Special Issue Paper

# AMReX and pyAMReX: Looking beyond the exascale computing project

**Andrew Myers[1], Weiqun Zhang[1], Ann Almgren[1], Thierry Antoun[1,2], John Bell[1], Axel Huebl[1] and Alexander Sinn[3]**

**Abstract**

AMReX is a software framework for the development of block-structured mesh applications with adaptive mesh refinement (AMR). AMReX was initially developed and supported by the AMReX Co-Design Center as part of the U.S. DOE Exascale Computing Project (ECP), and is continuing to grow post-ECP. In addition to adding new functionality and performance improvements to the core AMReX framework, we have also developed a Python binding, pyAMReX, that provides a bridge between AMReX-based application codes and the data science ecosystem. pyAMReX provides zero-copy application GPU data access for AI/ML, in situ analysis and application coupling, and enables rapid, massively parallel prototyping. In this paper we review the overall functionality of AMReX and pyAMReX, focusing on new developments, new functionality, and optimizations of key operations. We also summarize capabilities of ECP projects that used AMReX and provide an overview of new, non-ECP applications.

## 1. Introduction

The AMReX software framework was initially developed as part of the U.S. DOE Exascale Computing Project (ECP) to support the development of block-structured adaptive mesh refinement (AMR) algorithms for multiphysics applications described by partial differential equations in simple or complex geometries (Zhang et al., 2021). Block-structured AMR uses a hierarchical representation of the solution at multiple levels of resolution where the solution on each level is defined on the union of data containers at that resolution. These data containers, which represent the solution over a logically rectangular subregion of the domain, can contain field data defined on a mesh, Lagrangian particles or combinations of both.

AMR reduces the computational cost and memory footprint compared to a uniform mesh while preserving accurate descriptions of different physical processes in complex multiphysics algorithms. AMReX builds on decades of development of the fundamental algorithmic underpinnings of adaptive mesh refinement, going back to work by Berger and Oliger (1984), Berger and Colella (1989) and Almgren et al. (1998). There are also a number of different open-source AMR frameworks; see the 2014 survey paper by Dubey et al. (2014) for a discussion of these frameworks and references to both the frameworks and application codes that use them.

The design of AMReX was informed by two major factors. First, AMReX should be able to support a wide range of multiphysics applications with different performance characteristics without imposing restrictions on how application developers construct their algorithms. Careful consideration was given to separating the design of the data structures and basic operations from the algorithms that use those data structures. AMReX uses a layered design that provides a rich set of flexible tools that can be used by a wide range of applications. This layered design allows application developers to interact with the software at several different levels of abstraction. Applications can simply use the AMReX data containers and iterators and none of the higher-level functionality, both in stand-alone AMReX-based codes and in non-AMReX based codes that selectively use AMReX

[1]Lawrence Berkeley National Laboratory, Berkeley, CA, USA
[2]CEA, Paris, France
[3]Deutsches Elektronen-Synchrotron (DESY), Hamburg, Germany

**Corresponding author:**
Andrew Myers, Applied Mathematics and Computational Research Division, Lawrence Berkeley National Laboratory, MS 50A-3111, Berkeley, CA 94720, USA.
Email: atmyers@lbl.gov

functionality. At an intermediate level of functionality, applications can use the data structures and iterators for single- and multi-level operations but retain complete control over the time evolution algorithm, that is, the ordering of algorithmic components at each level and across levels. AMReX also provides developers the option to exploit additional functionality in AMReX that is designed specifically for AMR algorithms that subcycle in time. This layer provides stubs for the necessary operations such as advancing the solution on a level, correcting coarse grid fluxes with time- and space-averaged fine grid fluxes, averaging data from fine to coarse and interpolating in both space and time from coarse to fine.

The other major factor influencing the AMReX design was performance portability. Current HPC architectures typically include some type of GPU accelerator. Achieving high performance on these architectures requires efficient utilization of the accelerator. However, different accelerators have different capabilities and programming models. To isolate applications from a particular architecture and programming model without sacrificing performance, AMReX introduced a lightweight abstraction layer that effectively hides the details of the architecture from the application. This layer provides constructs that allow the user to specify what operations they want to perform on a block of data without specifying how those operations are carried out. AMReX then maps those operations onto the target hardware at compile time so that the hardware is utilized effectively. AMReX currently supports CUDA, HIP and SYCL for GPU acceleration and OpenMP for multi-core CPU architectures. AMReX-based applications have run successfully at scale on some of the largest supercomputers in the world, each of which utilizes quite different hardware: OLCF's AMD MI250X-based Frontier, NERSC's NVIDIA A100 machine Perlmutter, ALCF's Aurora, which uses Intel Xe GPUs, and Riken's Fugaku platform, which uses ARM A64FX CPUs, to name a few.

AMReX was successfully used by six of the ECP applications to meet their performance/capability goals. These applications represented a wide range of disciplines and computational paradigms. The particle-in-cell code WarpX (Fedeli et al., 2022; Vay et al., 2018) is used for simulations of particle accelerators, beams and laser-plasmas. The astrophysics code, Castro (Almgren et al., 2010, 2020), uses high-fidelity explicit algorithms for compressible flow in combination with self-gravity and nuclear reaction networks. The cosmology code, Nyx (Almgren et al., 2013), models compressible flow in an expanding universe with self-gravity and a Lagrangian particle representation of dark matter. The MFiX-Exa code (Musser et al., 2022; Porcu et al., 2023) models multiphase particle-laden flows with reactions and heat transfer effects in complex geometries. The combustion codes, PeleC (De Frahan et al., 2023), based on a fully compressible formulation, and PeleLMeX (Esclapez et al., 2023), based on a low Mach number formulation, model reacting flows with detailed kinetics and transport in complex geometries. The AMR-Wind code solves incompressible flow with additional physics to model atmospheric boundary layers, with coupling to an unstructured flow solver to model flow around turbine blades (Sharma et al., 2024).

In addition to the ECP applications, AMReX is also being used in applications ranging from beam dynamics (ImpactX: (Huebl et al., 2022a)) and plasma wakefield acceleration (HiPACE++: (Diederichs et al., 2022)) to geophysical fluid dynamics (ERF (Almgren et al., 2023) for atmospheres and REMORA, 2023 for oceans) to epidemiology (ExaEpi, 2023), to name a few. These new applications, in combination with performance improvements and the addition of new features for existing applications, are driving the addition of new capabilities to AMReX.

In this paper, we focus on capabilities that have been developed in the past 4 years. We first provide an brief overview of AMReX. We then discuss a number of enhancements to AMReX aimed at enabling new application codes and improving performance, particularly on GPU-accelerated architectures. Next, we discuss the python binding, pyAMReX, that provides access to AMReX data structures and operations from Python. Finally, we present some conclusions and discuss future directions.

## 2. AMReX overview

In this section, we provide an overview of the basic capabilities of AMReX; we refer the interested reader to Zhang et al. (2021) for details.

AMReX provides a set of data containers that can be used to store structured mesh and particle data on distributed memory systems. Although the focus is on supporting block-structured AMR codes, AMReX's flexible data structures and functions for operating on these data are also useful for building non-AMR applications.

The multi-dimensional array is one of the most fundamental data structures used by many scientific computing applications. AMReX is written in C++17, which does not have native multi-dimensional array support. `amrex::FArrayBox` is an AMReX class for storing multi-dimensional mesh data, with a Fortran-like syntax for accessing the data. It holds the data for a logically rectangular spatial region with its integer index bounds specified by `amrex::Box`. The framework supports data that is cell-centered, face-centered, edge-centered or nodal.

In AMReX, the spatial index is global, thus the lower bounds of the array in `FArrayBox` do not need to be zero. In fact, they can be negative, which is useful for applications solving partial differential equations. For block-structured AMR, the computational domain consists of multiple logically rectangular subdomains that are distributed among MPI processes. `amrex::MultiFab` is a container for a collection of `FArrayBox` objects that are locally owned. The `amrex::MultiFab` also contains metadata for communication between MPI processes. The `amrex::MFIter` class iterates over a `amrex::MultiFab` object in parallel, with each MPI process iterating over the FArrayBox objects it owns. amrex::MFIter also supports a logical tiling option for cache-blocking and multi-threaded executation with OpenMP.

AMReX also provides a amrex::ParIter class for iterating over distributed particle data. Particles in AMReX are associated with a grid and a level. (We note that even in the absence of mesh data we do construct a mesh hierarchy to organize the particles.) In general, particle data can be stored in Array-of-Struct (AoS) style, Struct-of-Array (SoA) style, or in some combination of the two. The original implementation of particles in AMReX used AoS only. Later, support for optional particle components stored in SoA style was added, but the limitation that the required particle data (positions and id numbers) be stored in AoS-style remained.

In addition to the basic data types, AMReX supports a multilevel embedded boundary representation of complex geometry; linear solvers for cell-centered and nodal data; asynchronous I/O in a native format readable by ParaView, VisIt and yt; mesh pruning; interfaces to hypre and PETSc solvers; and level set functionality for particle-wall interactions. In addition, the capability to use the bittree (Dhruv, 2024) algorithm for regridding has been added recently.

As noted above, AMReX uses MPI for communication between processes. Common communication patterns such as ghost cell exchange, copying data from one distribution to another, and communication between AMR levels (including interpolation, restriction, etc.) are created automatically and cached for performance. AMReX also provides MPI functionality for common global reduction operations across MPI ranks. AMReX aggregates messages into communication buffers to reduce latency.

## 3. New features and optimizations

In this section, we discuss new capabilities added to AMReX to improve application performance and to support new application codes. We particularly focus on features added to support and improve the performance of post-ECP applications, including ImpactX, ERF, ExaEpi and REMORA, and on supporting the Python interfaces described below. The impact of any specific optimization is application-dependent; however, the performance improvements can be significant. For example, kernel fusing provides significant improvement on GPUs for applications with many small boxes while particle intensive applications can see dramatic improvement from the new pure structure-or-arrays capability.

### 3.1. Performance portability

Many modern supercomputers are hybrid computing systems built with both CPUs and GPUs, with the latter providing most of the computing power. There are several different GPU vendors and their GPUs have different architectures. Programming for these heterogeneous systems is challenging because there is no standard programming model for all of these different GPU architectures. To facilitate heterogeneous programming for AMReX applications and for our own development of AMReX, we have created a performance portability abstraction layer, which is built upon the native solutions provided by the vendors. That is, we use CUDA, HIP and SYCL for NVIDIA, AMD and Intel GPUs, respectively. AMReX's portability layer is similar to those provided by Kokkos (Edwards et al., 2014), Alpaka (Zenker et al., 2016) or RAJA (Beckingsale et al., 2019), but tailored to the needs of block-structured AMR applications.

AMReX provides a construct, amrex::ParallelFor, for launching a GPU kernel looping over a one- or multi-dimensional iteration space. Listing 1 shows an example of using ParallelFor on three MultiFab objects. The example code runs on CPUs and all major GPUs. On multicore CPU architectures, we decompose the loops over the individual FArrayBox's into logical tiles to enable OpenMP parallelism and improve cache performance. The tile size is specified at runtime in the inputs file and can also be set to different values for different loops.

```
1  // triad on MultiFabs: mfa = mfb + scalar*mfc
2  #pragma omp parallel if (Gpu::notInLaunchRegion())
3  for (MFIter mfi(mfa, TilingIfNotGPU());
4       mfi.isValid(); ++mfi)
5  {
6    auto const& a = mfa.array(mfi);
7    auto const& b = mfb.const_array(mfi);
8    auto const& c = mfc.const_array(mfi);
9    ParallelFor(mfi.tilebox(),
10     [=] AMREX_GPU_DEVICE (int i, int j, int k) {
11       a(i,j,k) = b(i,j,k) + scalar * c(i,j,k);
12   });
13 }
```

Listing 1: Example of ParallelFor. This code can be compiled to run on CPU with OpenMP or GPU with CUDA, HIP, or SYCL.

Here, the data type of a, b and c is amrex::Array4, which is a lightweight alias to FArrayBox. Unlike FArrayBox, this non-owning data structure can be used in GPU kernels. The separation of data ownership and access is a very important design pattern for GPU programming. For these types of relatively simple kernels, high efficiency can be achieved. For example, the triad kernel in this example is memory bandwidth limited and is able to reach more than 80% of the peak memory bandwidth on NVIDIA A100 GPUs, despite the cost associated with multi-dimensional arrays such as the conversion between 1D and 3D indices.

## 3.2. Kernel fusing

AMR applications can have many relatively small patches when the regions identified as requiring higher resolution are small and cannot be merged to form larger patches. To launch a GPU kernel for each of the small patches would incur a significant cost due to GPU kernel launch latency. For example, a simple kernel like the triad in Listing 1 running on an AMD MI250X will only be able to achieve $\sim 10\%$ of its peak memory bandwidth on small boxes of $32^3$ cells. To alleviate the latency issue of launching multiple small kernels, AMReX provides a kernel fusion capability. Listing 2 shows an example of rewriting the code in Listing 1 to use a single kernel launch for all the local data in the `MultiFab`.

```
// triad on MultiFabs: mfa = mfb + scalar*mfc
auto const& a = mfa.arrays();
auto const& b = mfb.const_arrays();
auto const& c = mfc.const_arrays();
ParallelFor(mfa,
  [=] AMREX_GPU_DEVICE (int box,
                        int i, int j, int k) {
    a[box](i,j,k) = b[box](i,j,k)
                  + scalar * c[box](i,j,k);
});
```

Listing 2: Example of kernel fusion. Only one GPU kernel is launched even when there are multiple patches. Note that in this new approach, the MFIter for loop is no longer needed due to kernel fusing inside this version of the ParallelFor function. This code can be compiled to run on CPU with OpenMP or GPU with CUDA, HIP, or SYCL.

In this example, if there are 512 patches of $32^3$ cells each, only one GPU kernel is launched to work on all 512 patches, which enables it to achieve similar performance as if operating on a single patch of $256^3$ cells. For tests with 512 batches of $32^3$ cells on one Graphics Computing Die (GCD) of an AMD MI250X, it takes about $1.16 \times 10^{-3}$ and $3.67 \times 10^{-4}$ seconds without and with kernel fusing, respectively. For tests with a single patch of $256^3$ cells, it takes about $3.28 \times 10^{-4}$ and $3.35 \times 10^{-4}$ seconds, for the old and new approaches, respectively. The new approach is much faster for cases with many small kernels, and it is only slightly slower for cases with only a couple of very large kernels. Also note that AMReX by default uses multiple GPU streams. If only one stream is used, the test with 512 small kernel launches would take about $1.93 \times 10^{-3}$ seconds.

## 3.3. Compile-time kernel specialization

The development of AMReX is often driven by the needs of AMReX-based applications. A number of these applications often use runtime parameters in GPU kernels such as shown in Listing 3, which shows an example with six possible options for execution with branches based on runtime conditions.

```
int A_runtime_option = ...;
int B_runtime_option = ...;
enum A_options : int { A0, A1, A2, A3 };
enum B_options : int { B0, B1 };
ParallelFor(N, [=] AMREX_GPU_DEVICE (int i)
{
    // ...
    if (A_runtime_option == A0) {
        // ...
    } else if (A_runtime_option == A1) {
        // ...
    } else if (A_runtime_option == A2) {
        // ...
    else {
        // ...
    }
    if (A_runtime_option != A3 &&
        B_runtime_option == B1) {
        // ...
    }
    // ...
});
```

Listing 3: Example of GPU kernel with runtime options.

Because the runtime options are the same for all threads, thread divergence is not an issue. However, different branches may place drastically different demands on computing resources such as registers. The compiler could not know at compile time which branch will be taken. Even if the cheapest branch is taken at run time, the performance could be severely affected by the assumption that the most expensive branch might be taken. To address this performance issue, AMReX provides a special version of the `ParallelFor` kernel-launching construct that is optimized for this type of case.

```
int A_runtime_option = ...;
int B_runtime_option = ...;
enum A_options : int { A0, A1, A2, A3 };
enum B_options : int { B0, B1 };
ParallelFor(
    TypeList<CompileTimeOptions<A0,A1,A2,A3>,
             CompileTimeOptions<B0,B1>
            >{},
    {A_runtime_option, B_runtime_option},
    N,
    [=] AMREX_GPU_DEVICE (int i,
                          auto A_control,
                          auto B_control) {
    // ...
    if constexpr (A_control.value == A0) {
        // ...
    } else if constexpr (A_control.value == A1) {
        // ...
    } else if constexpr (A_control.value == A2) {
        // ...
    else {
        // ...
    }
    if constexpr (A_control.value != A3 &&
                  B_control.value == B1) {
        // ...
    }
    // ...
});
```

Listing 4: Example of the optimized version of a GPU kernel with runtime options.

Using this form of `ParallelFor`, the example in Listing 3 can be rewritten as shown in Listing 4. Here, the code is expanded into all combinations of the run time parameters at compile time using C++17 fold expressions. The resulting code does not incur an unnecessary performance penalty if the expensive branches are not taken.

This feature was first developed when we were optimizing WarpX's `GatherAndPush` kernel, which has the runtime option of using a QED physics module or adding external fields evaluated with an expensive mathematical expression parser. Without the kernel specialization, the `GatherAndPush` kernel in a typical WarpX benchmark without using QED or external fields at runtime took about 2.17 s on 1 GCD of AMD MI250X. It reduced to 1.34 s with the kernel specialization due to double occupancy.

## 3.4. Parallel reductions

AMReX provides a flexible and performance portable approach for common node-level reductions used by applications. Listing 5 shows an example of computing the sum, minimum and maximum of a `MultiFab` on a parallel compute device (GPU or multi-core CPU). As mentioned earlier, global reduction primitives for multi-node reductions via MPI are also provided.

```
1  auto const& a = mf.const_arrays();
2  auto result = ParReduce(
3    TypeList<ReduceOpSum,ReduceOpMin,ReduceOpMax>{},
4    TypeList<Real,Real,Real>{},
5    mf,
6    [=] AMREX_GPU_DEVICE (int box,
7                          int i, int j, int k)
8    {
9      auto v = a[box](i,j,k);
10     return {v, v, v};
11   }
12 );
13 // result is a tuple holding the sum, min and max.
```

Listing 5: Performing an on-device, parallel reduce on mixed reduction types.

```
1  ReduceOps<ReduceOpSum,ReduceOpMin,ReduceOpMax> ops;
2  using reduce_data = ReduceData<Real, Real, int>;
3  auto r = ParticleReduce<reduce_data> (
4      pc, [=] AMREX_GPU_DEVICE (const PType& p)
5      -> GpuTuple<Real, Real, int>
6      {
7          // sum over the first real component
8          const Real a = p.rdata(0);
9          // min over the second real component
10         const Real b = p.rdata(1);
11         // max over the first integer component
12         const int c = p.idata(0);
13         return {a, b, c};
14 }, ops);
15 // result is a tuple holding the sum, min and max.
```

Listing 6: Performing an on-device, parallel reduce on mixed reduction types for particle data. The argument 'pc' is an AMReX particle container, and 'PType' is the type of particles it contains. The above code will work for both Array-of-Struct and pure Struct-of-Array data (see below for more detail).

These parallel reduction operations for mixed data types also work for particle data, as shown in Listing 6. This is useful, e.g., in the AMReX-based epidemiological modeling code ExaEpi. Every day, the total number of infected, hospitalizations, people on ventilators, etc. Need to be tracked for diagnostic purposes. Fusing these global reduction operations minimizes the number of GPU kernel launches needed.

## 3.5. MPI communication of mesh and particle data

We have added support to AMReX for GPU-aware MPI. If it is enabled, AMReX allocates the buffers from a dedicated memory arena to improve performance. Moving data between the buffers and data containers (e.g., `MultiFab`) could be costly if care is not taken to minimize the GPU kernel launch overhead. When packing/unpacking the buffer, we often need to copy data from/to hundreds of small regions, which could be as small as a single cell. We use kernel fusion to ameliorate this cost. For mesh data, we have also added support for function templates for user-defined index mapping to capture application-specific communication patterns, which can be useful for non-Cartesian coordinates and multiblock applications, which will be discussed further in the section on future work.

## 3.6. Pure struct-of-array particles

Particles in AMReX are associated with a level and grid in the adaptive mesh refinement hierarchy based on their position coordinates. For CPU-only applications, grids are typically further subdivided into tiles for OpenMP threading and efficient cache-blocking (Zhang et al., 2021).

As a result, particles in AMReX are required to have `AMREX_SPACEDIM` position components. In addition, a 64-bit unsigned integer id number that uniquely identifies each particle both on and across all MPI ranks is also required; currently, AMReX supports creating up to $\sim 5 \times 10^{11}$ unique particles on each of up to $\sim 16.7 \times 10^{6}$ MPI ranks ($\sim 9 \times 10^{18}$ total unique particles). These required components can be extended by an arbitrary number of real- and integer-valued optional components that can be used to store application-specific data (e.g., mass, momentum, moment of inertia, ionization level, etc.). These optional components can either be registered at compile time or added at run time.

As discussed earlier, AMReX initially required the particle coordinates and id number to be stored in an AoS-style. This layout is non-optimal in several ways. First, any time the particle positions were accessed, the id numbers for those particles would also be read into cache. If the id numbers (mostly used for postprocessing and to flag particles for later removal) were not needed, between 1/4 (double precision floats, 3 spatial dimensions) and 2/3 (single precision floats, 1 spatial dimension) of the available memory bandwidth was wasted reading in data that wouldn't be used. A similar point could made for operations that only need to access and/or modify the id number and not the particle position coordinates. Second, the AoS layout is unable to take advantage of SIMD instructions (either manual or auto-vectorization), in which the exact same operations are applied to adjacent (and properly aligned) memory locations. Taking advantage of these instructions is crucial to achieving maximum performance on modern CPU architectures. An SoA data layout can also better take advantage of vectorized loads and stores, in which consecutive threads access consecutive memory locations, on GPU architectures as well.

To address these limitations, we have added to AMReX the ability to store particle data, including the position and id components, in pure SoA style. Figure 1 illustrates the differences in data layout for the original AMReX implementation where position and id numbers were stored in AoS form and the new pure SoA-style. Our implementation of this feature makes heavy use of template metaprogramming with C++17 features such as `constexpr if` to maintain compatibility with the legacy AoS-style particle layout and to ensure that migrating codes from the old to the new layout is minimally invasive. Listing 7 demonstrates how to use the new capabilities. The code at the top is AoS-only, while the code on the bottom works for both AoS-style and SoA-style particle data. Several run-able examples that demonstrate how to use this new feature are available online.[1,2,3]

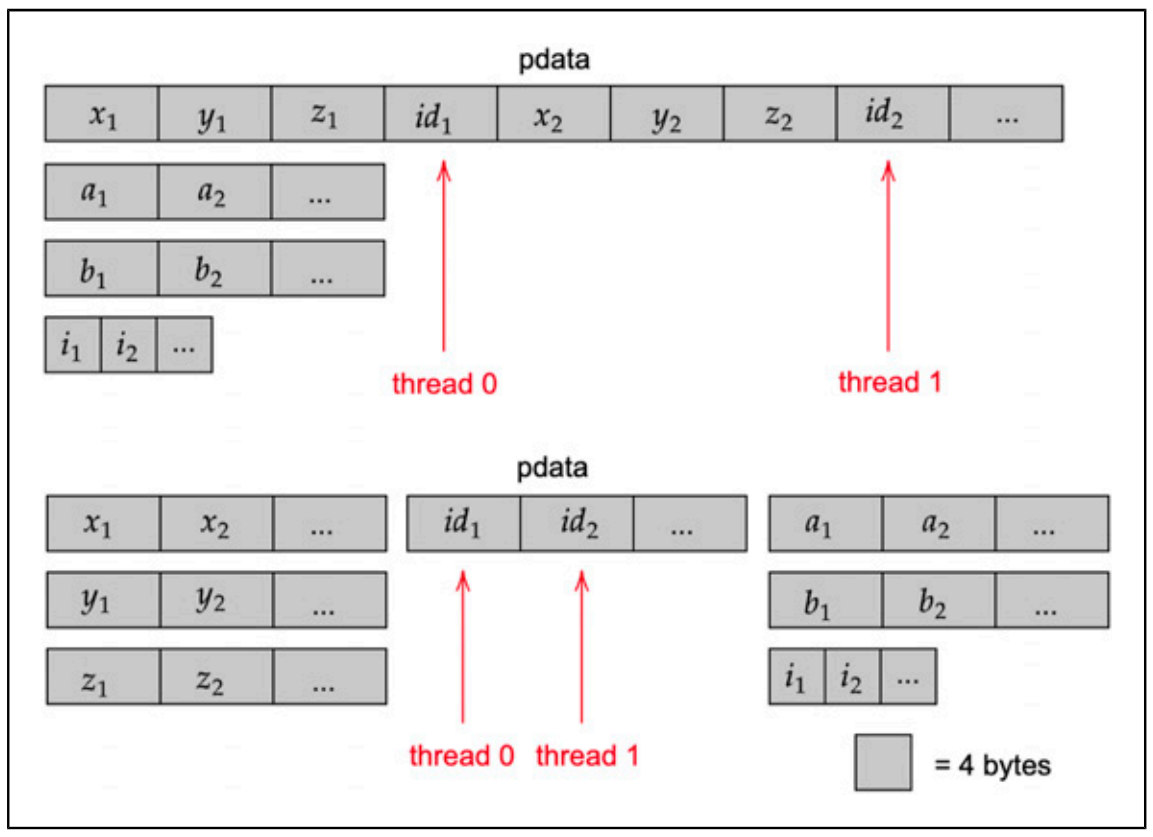

**Figure 1. Top:** Old particle data layout with particle positions *x*, *y*, and *z* and the unique *id* number stored AoS-style, along with three additional components, *a*, *b*, and *i* stored SoA-style, where *a* and *b* are real-valued and *i* is integer-valued. This example shows the case of 3 spatial dimensions and double precision floating point data. **Bottom:** New particle data layout where all quantities are stored SoA-style. Note that, for GPU execution, consecutive threads access adjacent memory locations when reading in a component (say, the *id* numbers) for the new layout but not the old. This enables vectorized loads and stores and does not waste memory bandwidth on data that will not be used. The new layout also enables better cache reuse and vectorization on CPU architectures as well.

Since these changes, several AMReX-based codes have made the transition to pure SoA particle data. In particular, we studied the effect of this change on the performance of ImpactX (Huebl et al., 2022a), a beam dynamics code including space charge effects that uses AMReX for its particle and mesh data structures as well as GPU offloading. ImpactX is the successor of the IMPACT-Z code for beam dynamics modeling in particle accelerators (Qiang et al., 2000).

```
// old style - positions and ids use AoS layout,
// radius uses SoA
auto& tile = pti.GetParticleTile();
const auto np = tile.numParticles();
auto pdata = tile.getParticleTileData();

amrex::ParallelFor(np,
[=] AMREX_GPU_DEVICE (int i)
{
    // read id and pos from the AoS
    const auto& p = pdata.m_aos[i];
    if (p.id().is_valid()) {
        Real x2 = (p.pos(0) - x0)*(p.pos(0) - x0);
        Real y2 = (p.pos(1) - y0)*(p.pos(1) - y0);
        Real z2 = (p.pos(2) - z0)*(p.pos(2) - z0);
        Real r = sqrt(x2 + y2 + z2);
        // write radius to component 0 of SoA
        pdata.rdata[0][i] = r;
    }
});
```

```
// new style - same code will work whether the
// positions and ids use AoS or SoA
auto& tile = pti.GetParticleTile();
const auto np = tile.numParticles();
auto pdata = tile.getParticleTileData();

amrex::ParallelFor(np,
[=] AMREX_GPU_DEVICE (int i)
{
    // wrapper that can handle AoS or SoA
    const auto& p = pdata[i];
    if (p.id().is_valid()) {
        Real x2 = (p.pos(0) - x0)*(p.pos(0) - x0);
        Real y2 = (p.pos(1) - y0)*(p.pos(1) - y0);
        Real z2 = (p.pos(2) - z0)*(p.pos(2) - z0);
        Real r = sqrt(x2 + y2 + z2);
        // write radius to component 0 of SoA
        pdata.rdata[0][i] = r;
    }
});
```

Listing 7: Example of migrating a ParIter loop from the old layout to the new one. In both cases, we loop over all the particles, compute a radius from their Cartesian position coordinates, and write the result to the Struct-of-Arrays. The first code snippet illustrates the old style in which the particle ids and positions were always stored in an AoS. The second snippet is coded in the new style, which will work for both pure SoA and legacy-style data layouts. The only difference between the two snippets is in line 11, which no longer makes any mention of an AoS. When compiled using the legacy data layout, the type of '`p`' will be a specialization of the '`amrex::Particle`' template, a simple struct. When compiled using the new, pure SoA layout, '`p`' will be a specialization of the '`amrex::SoAParticle`', a wrapper that knows how to access the underlying data regardless of its layout. The '`auto`' allows the same code to work for both.

Table 1 summarizes the improvement benefits seen in ImpactX from transitioning its particles to the new, pure SoA format. As a benchmark, we use a standard test problem that propagates a particle beam through a single FODO cell, that is through one cell of lattice constructed of alternating focusing and defocusing quadrapolar lenses. The performance improvements are shown for a few select routines and also for the overall runtime of the test. For the GPU test, we used a single NVIDIA A100 SMX with 80 GB of HBM. The CPU tests were run on a single core of a 12th Gen Intel(R) Core(TM) i9-12900H. Both single precision (SP) and double precision (DP) results are shown. Overall, most routines see at least a small improvement from the change in data structure and bandwidth savings, with improvements of 2-3x not being uncommon due to synergies with improved GPU occupancy, compiler auto-vectorization for simple functions, and overhauled particle ID functions. Finally, as an added benefit, the new layout also enables the zero-copy GPU Python interface described in a later section for the particle position and id numbers.

**Table 1.** ImpactX speedup from transition to pure SoA data layout and optimized particle id handling. DP: double precision, SP: single precision.

| | GPU | | CPU | |
|---|---|---|---|---|
| | SP | DP | SP | DP |
| Push::Drift | 2.32 | 1.80 | 2.17 | 1.76 |
| Push::Quad | 2.32 | 1.32 | 1.05 | 1.00 |
| collect_lost | 2.78 | 3.65 | 2.48 | 3.24 |
| add_particles | 1.05 | 1.06 | 1.02 | 1.02 |
| Overall | 2.34 | 1.65 | 1.28 | 1.21 |

### 3.7. Memory management

On heterogeneous systems, there are a variety of memory types: device memory, managed memory, pinned host memory, and pageable host memory. AMReX provides a number of memory allocators so that the user can choose the type of memory for allocation. To minimize the cost of data movement, AMReX by default allocates data in device memory, and a runtime parameter is provided for switching to managed memory. Allocating and deallocating GPU memory can be very costly. To minimize memory allocation, AMReX pre-allocates a large chunk of device memory during initialization and stores it in a memory arena. This significantly improves the performance for operations where temporary memory space is needed for example, in our implementation of computing the sum of the data in a `MultiFab`. It takes 0.0001 s on NVIDIA A100 for a `MultiFab` with $256^3$ cells using the memory arena; Without the memory arena, the cost is 0.0006 s.

A common mistake in GPU programming allows the memory used in a GPU kernel to be freed before the kernel finishes execution, due to the asynchronous nature of GPU kernel launches. One could add a synchronization call after the GPU kernel launch, but this could add unnecessary synchronization to the code. In AMReX, we have implemented an asynchronous-safe memory arena that does not require synchronization. An example is shown in Listing 8, in which the code is asynchronous-safe because we use `amrex::The_Async_Arena`.

```
{
  FArrayBox tmp(..., The_Async_Arena());
  ParallelFor(...);  // Async kernel using tmp
} // Scope ends: the destructor of tmp is called.
```

Listing 8: Example of using the asynchronous-safe memory arena.

The implementation for CUDA uses a stream ordered memory allocator that guarantees the free happens after the kernel finishes. For HIP and SYCL, we use a host callback function to implement the memory safety guarantee.

These memory arenas also work with particle data. In fact, with the inherently dynamic data structures necessitated by particle methods, the use of the arenas is particularly desirable. The memory arenas also allow flexibility in placing particle data in different types of memory for GPU execution. By calling the appropriate arena, users can place data in device memory, pinned host memory, managed memory, etc, based on how they plan to use this data in their application workflow. For example, for off-line or for in situ analysis, it often makes sense to perform the analysis on the host side, to avoid over-subscribing device memory. Both WarpX and the AMReX-based biological cell modeling code BoltzmannMFX (Palmer et al., 2023) frequently copy data into temporary host-side particle containers that use pinned memory that allows direct memory writes from GPU kernels, avoiding temporary device memory spikes.

## 4. pyAMReX

AMReX has seen wide adoption as a software library used in pre-compiled applications. With the popularity of interactive computing in recent years, for example, via Project Jupyter (Kluyver et al., 2016), and the rise of scripting languages as productive runtime "glue code" to connect and steer software components, using AMReX from a scripting layer for tasks like integration, data processing and/or prototyping is desirable.

The new pyAMReX project is part of the AMReX software ecosystem and builds directly on the AMReX C++ library. The Python language bindings of pyAMReX bridge the compute role of AMReX in block-structured simulation codes with the data science software ecosystem. With pyAMReX providing zero-copy GPU data access for workflows in AI/ML and in situ analysis, applications can be coupled and extended from Python, and rapid, massively parallel prototyping is enabled.

pyAMReX is implemented as a C++ library via pybind11 (Jakob et al., 2017), which targets the Python C API to expose standardized C entry points of Python modules. Due to the evolution of the C++ language from C++11 to C++17, type introspection and meta-programming inside pybind11 make adding new bindings for AMReX functions, classes, etc. Simple and maintainable.

```cpp
#include "pyAMReX.H"
#include <AMReX_Box.H>

// ...
    py::class_< Box >(m, "Box")
      // constructor with two arguments
      .def(py::init<  // C++ argument types
             IntVect const &,
             IntVect const &
           >(),
           // Python argument names (optional)
           py::arg("small"),
           py::arg("big")
      )
      // ...

      // method with two arguments
      .def("make_slab",      // Python name
           &Box::makeSlab,   // C++ function pointer
           py::arg("direction"),
           py::arg("slab_index")
           // Python doc string (optional)
           "Flatten the box in one direction."
      )
      // property (no arguments)
      .def_property_readonly(
        "ok",      // Python name
        &Box::ok  // C++ function pointer
      )
    ;
```

```python
import amrex.space3d as amr

small_end = amr.IntVect()
big_end = amr.IntVect(2, 3, 4)
b = amr.Box(small=small_end, big=big_end)

b                   # <amrex.Box of size '(3,4,5)'>
b.ok                # True
b.make_slab(1, 1)   # <amrex.Box of size '(3,1,5)'>
```

Listing 9: pyAMReX uses plain C++ code to register Python types. The top snippet defines the AMReX class Box in Python with pybind11, with a Python class constructor, a method and a property. The lower snippet shows the usage from Python.

Listing 9 shows how to expose the C++ type amrex::Box to Python; once it is registered other classes and methods can use this type in their Python interfaces. Following the same logic, applications that are built on AMReX just need to add minimal Python bindings themselves, because they can rely on all the Python-registered AMReX types from pyAMReX.

Python modules written with pybind11 are precompiled to shared libraries, which automatically expose only a set of standardized, C API symbols for import into the Python interpreter, that ultimately call AMReX C++ routines. Additional type extensions that are more effectively written in pure Python are added at import time. In the future, pyAMReX might transition to nanobind (Jakob, 2022), which is a recent re-implementation of pybind11 APIs with improved compile-, link- and runtime of the binding code.

### *4.1. Zero-copy APIs*

Exchanging large, heap/GPU-allocated data across API interfaces in Python, without the need to create copies, benefits from the permissive memory access rights in operating systems, since the Python interpreter and all its dynamically loaded (imported) libraries reside in the same process. Thus, any process-allocated memory can be accessed across Python modules and loaded C++ libraries by exchanging a pointer and the respective meta information that describe data placement and layout.

Luckily, in the past years the exchange of multi-dimensional (ND) array data in Python has been standardized in CPU code via the array protocol (Consortium for Python Data API Standards, 2021) driven by NumPy (Harris et al., 2020), based on the earlier buffer protocol implementations in Python. Adding support for sharing a Python type's data with read/write access requires implementation of a __array_interface__ dictionary property for AMReX Python types, which describes primarily shape, stride, data types, description and data pointer, among other details. The standardized __cuda_array_interface__ uses nearly the same metadata, with an optional CUDA stream added. Since unified memory pointers are available in CUDA, a pointer to GPU data can be exchanged between AMReX and CuPy (Okuta et al., 2017), Numba (Lam et al., 2015), or PyTorch (Paszke et al., 2019) to create views into their respective host-controlled GPU data container that works the same way as it works with NumPy CPU array data.

A new iteration on the CPU/GPU array interfaces was pursued with the deep learning pack standard (DLPack). DLPack uniformly wraps CPU or GPU data and brings improvements in supported accelerated programming models, more supported packages in the data science ecosystem, and in lifetime (data ownership) handling. A DLPack implementation in pyAMReX is ongoing[4] and will enable ROCm and SYCL GPU support.

Listing 10 shows how the array interface is used in practice. Line 7 creates a CuPy ND array that is a readable and writable data view into an existing AMReX Array4. CuPy reads the Array4.__cuda_array_interface__ dictionary and extracts data pointer and layout information to create a non-owning cupy.ndarray. Equivalently, NumPy ND arrays and PyTorch tensors are created.

```python
import amrex.space3d as amr
import cupy as cp

mfab = amr.MultiFab(<...>)
for mfi in mfab:
    array4 = mfab.array(mfi)
    cp_arr = cp.array(array4, copy=False)
```

Listing 10: Standardized interfaces on Python types can be used to create views from one datatype into another. Here, a CuPy ND array is used to read-write access the field data in an AMReX Array4.

Convenient methods are added to types like Array4, for example, .to_numpy () and.to_cupy (), which spare the user from the extra line and also keep the ND index order in AMReX-typical Fortran order (indexed x,y,z,component as F_CONTIGUOUS, although C_CONTIGUOUS is the default for Python). Notably, in 2D benchmarks with CuPy, Fortran indexing caused runtime overheads of 44 %.[5] Thus, experienced users have the choice to set a.to_cupy (order = “C”) argument to index as C-ordered component,z,y,x and perform optimally with CuPy.

Additional arguments such as.to_numpy (copy = True) are supported and perform the expected device-to-host copy if called on a device memory container, and a device-to-device copy if performed as.to_cupy (copy = True). For users that use the managed memory arena in AMReX, it is worth noting that.to_numpy (copy = False) works for read and write on GPU memory as expected, but includes the performance penalty of implicit data transfers. As a common shorthand notation from CuPy for writing CPU/GPU agnostic code, .to_xp () methods are provided as well, which pick NumPy (np) or CuPy (cp) automatically (xp).

## 4.2. Computing in python

Currently, pyAMReX does not expose amrex::ParallelFor/Reduce/\ldots computing primitives, but can call pre-compiled functions from Python that make use of those. In order to be able to generate new compute kernels at runtime from the Python scripting language, pyAMReX makes use of the aforementioned zero-copy APIs to pass data for compute into a third-party Python CPU/GPU framework of choice, for example, NumPy, CuPy, Numba, PyTorch, which then can support just-in-time (JIT) GPU kernel compilation.

Listing 11 shows how to generate accelerated array computations on an ND field (AMReX MultiFab). As is typical in AMReX, the outermost loops iterate over mesh-refinement levels and data blocks on the current device. Line 21 creates a CuPy or NumPy array view, followed by a device-side assignment of the value 42.0 to all index points using the () ellipsis, independent of array rank.

```python
# finest active MR level, get from a
# simulation's AmrMesh object, e.g.:
# finest_level = sim.finest_level

# iterate over mesh-refinement levels
for lev in range(finest_level + 1):
    # get an existing MultiFab, e.g.,
    # from a simulation:
    # mfab = sim.get_field(lev=lev)

    # grow (aka guard/ghost/halo) regions
    ngv = mfab.n_grow_vect

    # get every local block of the field
    for mfi in mfab:
        # global index box w/ guards
        bx = mfi.tilebox().grow(ngv)

        # NumPy/CuPy representation: non-
        # copying view, w/ guard/ghost
        field =  mfab.array(mfi).to_xp()

        field[()] = 42.0
```

Listing 11: An example Python script showing accelerated array operations on a field with `pyAMReX` v24.05.

One difference when operating on NumPy/CuPy views compared to AMReX Array4 is the index offset of each local data block. The former use zero-based indexing on local blocks while AMReX uses global indexing.

In Listing 11, only the select-all array operation [()] = 42.0 represents the “hot loop” of operations (or on GPU, the kernel function). Other Python array ellipsis selectors such as operating on an index range with an offset (e.g., [1:,1:, 1:]) or a slice (e.g., [0, \ldots ]), etc. Can be used as usual to form more complex expressions such as stencils. In frameworks like CuPy and Numba, multiple subsequent array operations can also be fused into one kernel, using their respective framework JIT syntax.

Particle operations follow the same logic, as shown in Listing 12. Both the legacy AoS + SoA and the new pure SoA particle layout are supported in pyAMReX. The pure SoA particle layout provides better performance as well as write access for GPU particle positions and indices in CuPy.[6]

```python
# code-specific getter function, e.g.:
# pc = sim.get_particles()

# iterate over mesh-refinement levels
for lvl in range(pc.finest_level + 1):
    # loop local tiles of particles
    for pti in pc.iterator(pc, level=lvl):
        # compile-time and runtime
        # attributes in NumPy/CuPy
        soa = pti.soa().to_xp()

        x = soa.real["x"]
        y = soa.real["y"]

        # write to all particles in tile
        x[:] = 0.30
        y[:] = 0.35
        soa.real["z"][:] = 0.40

        soa.real["a"][:] = x[:]**2
        soa.real["b"][:] = x[:] + y[:]
        # ...

        # all int attributes
        for soa_int in soa.int.values():
            soa_int[:] = 12
```

Listing 12: An example Python script showing accelerated particle operations with `pyAMReX` v24.05.

As for the ND field arrays, the “hot loop” of computation is best placed in array operations such as [:] = value. Selective indexing and masking with temporary arrays are the usual patterns in high-performance Python, and various Python frameworks support JIT compilation of explicit indices when needed.

### 4.3. Project integration

pyAMReX was initially designed with two workflows in mind: (i) to enhance an existing AMReX application with Python code, data-science and/or AI/ML capabilities, (ii) to write a standalone application or test of AMReX, rapidly prototyped in Python.

Enhancing, steering and modifying HPC applications with routines written in Python is a desirable feature in exascale-ready codes such as WarpX, which already benefited from a runtime modularity in the (CPU-only) predecessor code WARP (Grote et al., 2005). Spearheading the adoption of pyAMReX for exascale modeling codes are two Beam, Plasma & Accelerator Simulation Toolkit (BLAST) codes (Huebl et al., 2022b) built on AMReX, WarpX and ImpactX, that now use user-defined C++ to Python callbacks and manipulate AMReX-managed GPU data via pyAMReX: users adjust the treatment of selected particle species, prototype new physical accelerator elements or replace numerical solvers, load external data, and steer simulation behavior in a way that would be too custom or complex to define in a key-value based inputs file syntax.

An integration with a machine learning (ML) framework has recently been reported in Sandberg et al. (2024). In this work, the ImpactX code (Huebl et al., 2022a) is augmented using pyAMReX to couple analytical particle beam updates and data-driven particle updates using a pre-trained neural network. The ML training data was created with a high-fidelity WarpX simulation and used to train a neural network offline. The resulting model is then integrated with the CUDA bindings of pyAMReX to update the GPU-accelerated ImpactX particle simulation state with GPU-accelerated neural network inference using PyTorch. For particle accelerator physicists, replacing traditionally costly sections of a beamline simulation with such specialized surrogate models enables the study of whole-device models in start-to-end simulations.

Finally, pyAMReX also supports the design of standalone benchmarks, prototype examples, new AMReX-powered command line tools and even applications, purely from Python. AMReX tutorials have been updated to include examples of the Python usage of the MultiFab class and to solve a multi-dimensional heat equation, providing easy accessibility of AMReX to new community members and for use in education. Additionally, being able to integrate AMReX functionality in scripts will enable users and developers to build standalone data processing tools, for example, for post-processing, and will enable rapid prototyping and unit testing of AMReX functionality and numerical solvers based on AMReX.

## 5. Conclusions and future work

In this paper, we have reviewed the basic features of AMReX and discussed a number of new features that have been added to the framework. One of these new features supports the aggregation of a number of small computational kernels into a single kernel launch. A related feature incorporates an efficient kernel launching mechanism that supports selection of different kernel options at runtime. We have also developed a flexible reduction mechanism that performs a number of different reductions simultaneously.

We have made several enhancements to the particle functionality in AMReX. We have introduced a more flexible approach for representing particle data that can combine both SoA and AoS representations. In particular, AMReX now supports a pure SoA representation that has led to significant performance improvements for a number of applications. We have also developed a Python binding of AMReX, pyAMReX, that provides a bridge between AMReX simulation methodology and data science software.

There are a number of additional capabilities that we are currently developing within AMReX to support the requirements of new applications. One area of active development is to extend interoperability using a multiblock paradigm to couple different applications efficiently. In a multiblock approach, each application has its own index space and the framework provides the mapping between index spaces and between (possibly) different solution representations. The applications can then be run in an MPMD mode using separate MPI commnicators with AMReX handling the communication between the different codes needed for the overall simulation. This facilitates coupling different applications without significant recoding. Use cases here range from solving a system of PDEs on a mapped multiblock domain that cannot be represented with a global index space to coupling different PDE solvers in different domains to coupling a stochastic algorithm such as kinetic Monte Carlo to a fluctuating hydrodynamics solver to model a catalytic surface interacting with a fluid. A related area of development is to support algorithm refinement approaches in which the representation of the underlying physics changes as a function of resolution. A hybrid algorithm that dynamically switches from a coarse-grained continuum model to a fine-grained particle model depending on flow conditions would be a typical use case for this capability. We are also extending the python interface to AMReX to support integration of ML concepts into AMReX-based applications. Use cases here would include developing ML-based surrogates for outer loop operations and developing ML methodologies for coupling between different scales.

Another driver for AMReX development is extending the performance portability constructs to work optimally on new architectures or new features in the software stack on existing architectures. One particularly interesting direction would be to use AMReX as the basis for the co-design of custom architectures. The developments of chiplet technology have the potential to support the design of performant, energy efficient specialized architectures for select applications.

## Acknowledgements

We acknowledge the AMReX open source community for their invaluable contributions. In addition, we acknowledge and thank Shreyas Ananthan, Ryan Sandberg and all other pyAMReX contributors. This research was supported by: the Exascale Computing Project (17-SC-20-SC), a collaborative effort of the U.S. Department of Energy Office of Science and the National Nuclear Security Administration; the U.S. Department of Energy, Office of Science, Office of Advanced Scientific Computing Research, Exascale Computing Project under contract DE-AC02-05CH11231; the U.S. Department of Energy (DOE) Office of Advanced Scientific Computing Research (ASCR) via the Scientific Discovery through Advanced Computing (SciDAC) program FASTMath Institute; and the Laboratory Directed Research and Development Program of Lawrence Berkeley National Laboratory under U.S. Department of Energy Contract No. DE-AC02-05CH11231. This research used resources of the National Energy Research Scientific Computing Center, a DOE Office of Science User Facility supported by the Office of Science of the U.S. Department of Energy under and resources of the Oak Ridge Leadership Computing Facility at the Oak Ridge National Laboratory, which is supported by the Office of Science of the U.S. Department of Energy under Contract No. DE-AC05-00OR22725.

## Declaration of conflicting interests

The author(s) declared no potential conflicts of interest with respect to the research, authorship, and/or publication of this article.

## Funding

The author(s) disclosed receipt of the following financial support for the research, authorship, and/or publication of this article: This work was supported by the U.S. Department of Energy, DE-AC02-05CH11231.

## ORCID iDs

Andrew Myers https://orcid.org/0000-0001-8427-8330
Ann Almgren https://orcid.org/0000-0003-2103-312X
Thierry Antoun https://orcid.org/0000-0002-1003-0698
John Bell https://orcid.org/0000-0002-5749-334X
Axel Huebl https://orcid.org/0000-0003-1943-7141

## Notes

1. https://github.com/AMReX-Codes/amrex/tree/development/Tests/Particles/SOAParticle
2. https://github.com/AMReX-Codes/amrex/tree/development/Tests/Particles/RedistributeSOA
3. https://github.com/AMReX-Codes/amrex/tree/development/Tests/Particles/SENSEI_Insitu_SOA
4. https://github.com/AMReX-Codes/pyamrex/issues/9
5. https://github.com/cupy/cupy/issues/7783
6. https://github.com/cupy/cupy/issues/2031

*Andrew Myers* is a Computer Systems Engineer in the Center for Computational Sciences and Engineering at Lawrence Berkeley National Laboratory. His work focuses on scalable particle methods for emerging architectures in the context of adaptive mesh refinement. He is an active developer of the AMReX framework and of the electromagnetic Particle-in-Cell code WarpX.

*Weiqun Zhang* is a Computer Systems Engineer at Lawrence Berkeley National Laboratory. He is interested in high-performance computing, computational physics, and programming in general. Currently, he works on the AMReX software framework and WarpX, an advanced electromagnetic Particle-in-Cell code.

*Ann Almgren* is a Senior Scientist at Lawrence Berkeley National Laboratory, and the Department Head of the Applied Mathematics Department in LBL’s Applied Mathematics and Computational Research Division. Her primary research interest is in computational algorithms for solving PDE’s in a variety of application areas. Her current projects include the development and implementation of new multiphysics algorithms in high-resolution adaptive mesh codes that are designed for the latest hybrid architectures. She is a Fellow of the Society of Industrial and Applied Mathematics and was the Deputy Director of the ECP AMReX Co-Design Center.

*Thierry Antoun* pursued his engineering studies at ENSTA Paris, followed by an MSc in Applied Mathematics, specializing in Modeling and Simulation at the Institut Polytechnique de Paris. He was a summer intern at Lawrence Berkeley National Laboratory, focusing on optimizing AMReX particle data structures and the PIC codes WarpX & ImpactX. Since the end of 2023, he is a member of the CExA team at the French Alternative Energies and Atomic Energy Commission (CEA), working on the Kokkos library to facilitate the transition of codes to GPUs.

*John Bell* is a Senior Scientist at Lawrence Berkeley National Laboratory. His research focuses on the development and analysis of numerical methods for partial differential equations arising in science and engineering. He is a Fellow of the Society of Industrial and Applied Mathematics and a member of the National Academy of Sciences, and was the Director of the ECP AMReX Co-Design Center.

*Axel Huebl* is a Research Scientist and computational physicist at Lawrence Berkeley National Laboratory. He works at the interfaces of HPC, laser-plasma physics, and advanced particle accelerator research. He drives open standards in his community, is the software architect of the Beam, Plasma & Accelerator Simulation Toolkit (BLAST) and is a lead-developer of Exascale simulation software such as WarpX (2022 Gordon Bell Prize Winner) and ImpactX. He is an active developer of AMReX and leads pyAMReX.

*Alexander Sinn* is an undergraduate student at Deutsches Elektronen-Synchrotron (DESY). His research interest are in high-performance computing and modeling of particle accelerators. He is one of the maintainers of the BLAST simulation software HiPACE++ for wakefield accelerator modeling.